\documentclass[preprint,nofootinbib,aps,amsmath,amssymb,tightenlines]{revtex4}
\usepackage{amsmath,amssymb,amsbsy,latexsym,graphicx}

\newcommand{\ScSc}{\scriptscriptstyle}

\newcommand{\pb}[1]{\rlap{\lower1.5ex\hbox{$\longleftarrow$}}{#1}}
\newcommand{\spb}[1]{\rlap{\lower1.0ex\hbox{$\leftarrow$}}{#1}}
\newcommand{\sdpb}[1]{\rlap{\lower1.0ex\hbox{$\scriptstyle \Leftarrow$}}{#1}}
\newcommand{\dif}{\mathrm{d}}

\newcommand{\areaform}{\, {}^{\ScSc 2}\!\epsilon}
\newcommand{\gA}{{}^\gamma\!A}

\newcommand{\gSigma}{{}^\gamma\Sigma}

\newcommand{\Io}{\mathring{I}}
\newcommand{\Lo}{\mathring{L}}

\newcommand{\half}{\frac{1}{2}}
\newcommand{\rRe}{\mathrm{Re}}  
\newcommand{\rIm}{\mathrm{Im}}

\newcommand{\Tr}{\mathrm{Tr}}
\newcommand{\R}{\mathbb{R}}
\newcommand{\Z}{\mathbb{Z}}
\newcommand{\C}{\mathbb{C}}

\begin{document}

\preprint{IGPG--05/11-6}
\title{
Quantum geometry and black hole entropy: inclusion of distortion and rotation
\footnote{
Text based on parallel talk given at the VI Mexican School on
Gravitation and Mathematical Physics: ``Approaches to Quantum Gravity'', 
held in Playa del Carmen, Mexico, in November of 2004.
To appear in the Proceedings.  Research reported here was done jointly
with Abhay Ashtekar and Chris Van Den Broeck \cite{aev}.
}
}
\author{J Engle}
\email{engle@gravity.psu.edu}
\affiliation{Institute for Gravitational Physics and Geometry,\\
Physics Department, Penn State, University Park, PA 16802, USA}

\begin{abstract}
Equilibrium states of black holes can be modelled by isolated horizons. 
If the intrinsic geometry is spherical, they are called type I while if it 
is axi-symmetric, they are called type II. The detailed theory of geometry of
\emph{quantum} type I horizons and the calculation of their entropy can be 
generalized to type II, thereby including arbitrary distortions and 
rotations. The leading term in entropy of large horizons is again given 
by 1/4th of the horizon area for the \emph{same} value of the 
Barbero-Immirzi parameter as in the type I case. Ideas and constructions 
underlying this extension are summarized.
\end{abstract}

\maketitle

\section{Introduction}

Since the work by Bekenstein, Hawking, \textit{et al.}~in the early seventies 
on black hole thermodynamics, proposals have been made of various sorts for a 
microscopic explanation of black hole entropy.  Prominent among these
is the loop quantum gravity calculation of black hole entropy published
in 2000 \cite{abk} (see also \cite{abck} and earlier work cited therein).
The present work \cite{aev} aims at an extension of this calculation: 
the previous calculation restricted itself to the case in which the 
intrinsic geometry of the black hole horizon is spherically symmetric 
-- the ``type I" case.
We will refer to this previous calculation as the ``type I" calculation.  
In the present work, we extend the calculation to the inclusion of 
rotation and distortion of the horizon compatible with
axisymmetry.  This is referred to as the ``type II" case.

Both of these calculations assume \textit{isolated horizon} boundary
conditions at the horizon.  We will not go over the
definition of isolated horizons here; it is sufficient to say that 
a horizon is called ``isolated" if its intrinsic geometry is 
time-independent. Physically speaking, then, an isolated horizon 
represents a ``black hole in equilibrium."  For further details, 
see for example \cite{abl},\cite{ak}.

\section{Classical phase space and tools}

A key tool used in the extension of the calculation are certain
multipoles defined for isolated horizons \cite{aepv}:
\begin{equation}
I_n + i L_n := - \int_S Y_{n,0}(\zeta) \Psi_2 \areaform 
\end{equation}
where $S$ is a cross-section of the isolated horizon,
$\areaform$ is the area $2$-form on $S$, and
$(\zeta,\varphi)$ are the unique coordinates on $S$ 
in which the metric takes the standard form
\begin{equation}
\label{stdform}
q = R^2\left(\frac{1}{f(\zeta)} \dif \zeta^2 + f(\zeta) \dif \varphi^2\right).
\end{equation}
$\zeta$ takes values in $[-1,1]$ and $\varphi$ takes values
in $\R/2\pi\Z$.

The most important property for us that these multipoles have
is that they completely determine the intrinsic geometry of the
horizon upto diffeomorphism.  The plan is to come up with operators 
in the quantum theory corresponding to the multipoles which we will then 
use to characterize the quantum ensemble for which we calculate the entropy.
The importance of the multipoles and area determining the intrinsic
geometry upto diffeomorphism is that they thus determine the
``macroscopic state" of the black hole and hence form a good set of 
observables for fully characterizing the ensemble.

The approach we are taking is to quantize the following phase space.
Our basic variables are the Ashtekar-Barbero variables 
$(\gA_a^i, \gSigma_{ab}^i)$ (throughout this presentation the conventions
in \cite{al} are used).
We have an internal boundary which is $S^2 \times \R$, and on this 
boundary we impose isolated horizon boundary conditions, and 
require this isolated horizon to be type II, have \textit{fixed multipoles}
$\Io_n$,$\Lo_n$ and fixed area $a_o$.
On this phase space, it turns out that one cannot simply use the naive
symplectic structure
\begin{equation}
\Omega(\delta_1,\delta_2)
=-\int_M \Tr(\delta_1 \gA \wedge \delta_2 \gSigma
-\delta_2 \gA \wedge \delta_1 \gSigma) 
= \int_M \omega(\delta_1,\delta_2)
\end{equation}
where $\omega$ is the symplectic current derived from the action of the 
theory.
This is because such a symplectic structure is not preserved 
under time evolution.
\begin{figure}
  \begin{center}
  \includegraphics[height=3.6cm]{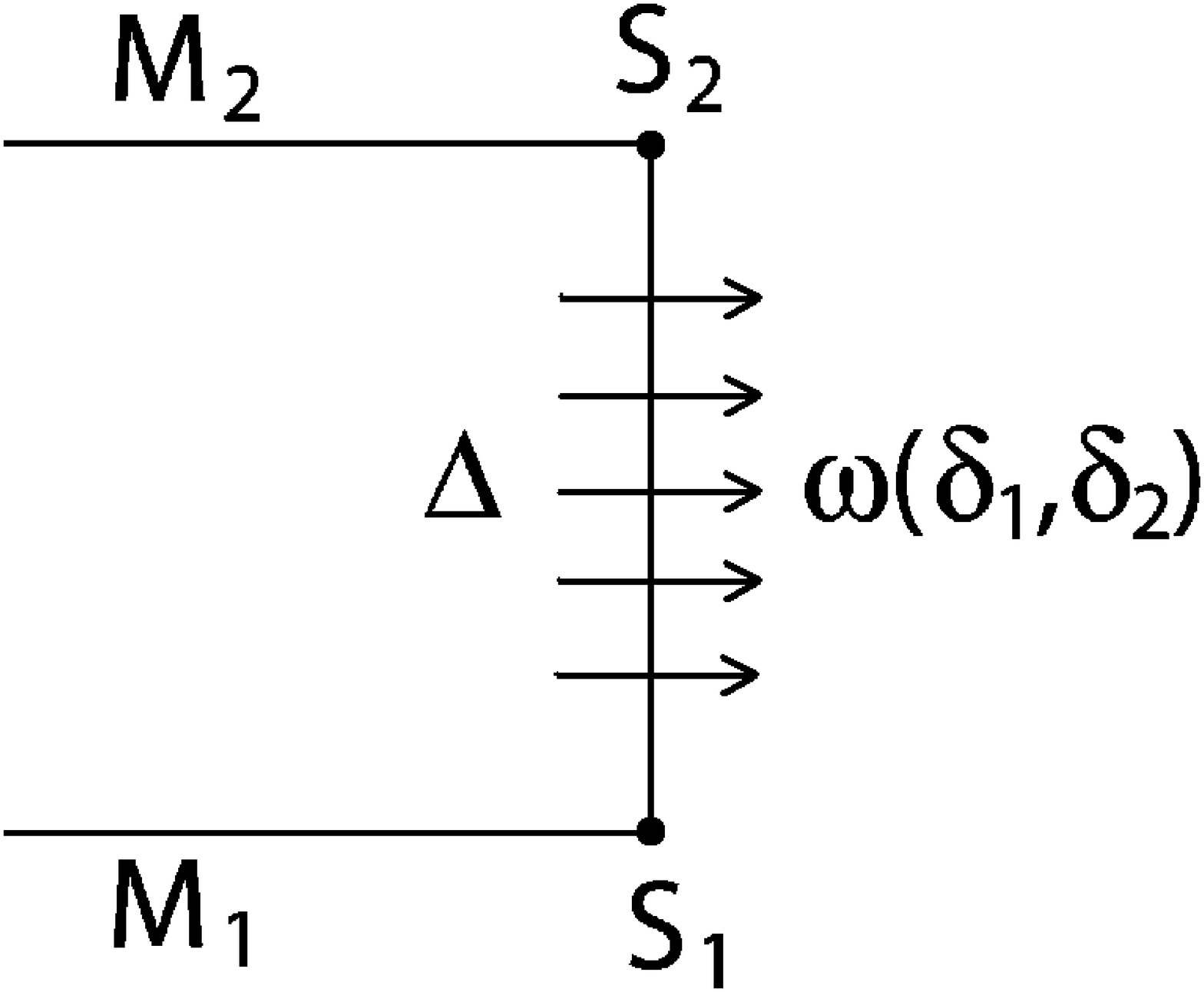}
  \caption{symplectic current escapes through the horizon}\label{sylcurrdiag}
  \end{center}
\end{figure}
Figure \ref{sylcurrdiag} shows the cause of this lack of preservation:
symplectic current is escaping 
across the horizon.  More precisely, we have
\begin{equation}
\label{fluxlost}
\int_{M_2} \omega(\delta_1,\delta_2) = 
\int_{M_1} \omega(\delta_1,\delta_2) 
- \int_{\Delta} \omega(\delta_1,\delta_2)
\end{equation}
The solution is to use the isolated horizon boundary conditions
to rewrite the $\Delta$-integral as
\begin{equation}
\label{rewriting}
\int_{\Delta}\omega(\delta_1,\delta_2)
=\left(\oint_{S_2}-\oint_{S_1}\right) \alpha(\delta_1,\delta_2)
\end{equation}
where $\alpha(\delta_1,\delta_2)$ is defined locally on $S_1$ and 
on $S_2$.  
(\ref{fluxlost}) then becomes
\begin{equation}
\int_{M_2}\omega + \oint_{S_2}\alpha = \int_{M_1}\omega + \oint_{S_1}\alpha
\end{equation}
so that
\begin{equation}
\Omega = \int_M \omega + \oint_S \alpha
\end{equation}
will work as a definition of the symplectic structure
that is invariant under time translations.
As one might suspect, there is an ambiguity in the choice
of $\alpha$.  Nevertheless, there is a natural resolution to this
ambiguity in the present context.

Before writing out explicitly the natural choice for $\alpha$, it will be
convenient to introduce two $U(1)$ connections on the horizon
$\Delta$.  
In our framework, as in \cite{abk}, for simplicity we fix an internal
vector field $r^i$ at the horizon and impose the partial gauge fixing
condition that $e^a_i r^i$ be the unit spatial normal to the horizon.
This reduces the gauge group at the horizon to $U(1)$.  The connections
${}^\gamma V$ and $W$ we are about to define represent connections on this
principal $U(1)$ subbundle. First we define
\begin{equation}
{}^{\gamma}V := \half \pb{\gA}^i r_i
\end{equation}
where the underarrow denotes pullback to $\Delta$.
In terms of ${}^{\gamma}V$, the boundary condition reflecting
that $\Delta$ is an isolated horizon takes the form
\begin{equation}
\label{quantumbc}
\dif {}^{\gamma}V = {}^\gamma \Psi_2 \areaform
= {}^{\gamma}\Psi_2 (8\pi\gamma)(\pb{\gSigma}\cdot r)
\end{equation}
where ${}^\gamma \Psi_2:= \rRe \Psi_2 + \gamma \rIm \Psi_2$.

We then define
\begin{equation}
W := {}^{\gamma}V + \frac{1}{4}(\mathring{f}'-f')\dif \varphi 
- \frac{\gamma}{2} \omega
\end{equation}
where $f$ and $\varphi$ are as in (\ref{stdform}), $\omega$ is the
rotation one-form on the horizon (see \cite{abl}), 
$\mathring{f}:=1-\zeta^2$, and the prime denotes derivative
with respect to $\zeta$.  In terms of $W$, the natural choice for $\alpha$
is
\begin{equation}
\label{alphaeq}
\alpha(\delta_1,\delta_2) 
= \frac{1}{8\pi G}\frac{a_o}{\gamma \pi} \delta_1 W \wedge \delta_2 W
\end{equation}
so that our symplectic structure is given by
\begin{equation}
\Omega(\delta_1,\delta_2)
=-\int_M \Tr(\delta_1 \gA \wedge \delta_2 \gSigma
-\delta_2 \gA \wedge \delta_1 \gSigma) 
+\frac{1}{8\pi G}\frac{a_o}{\gamma \pi} \oint_S \delta_1 W \wedge \delta_2 W
\end{equation}

It is also important to note that, in terms of $W$, the boundary condition
(\ref{quantumbc}) is now equivalent to a condition of the familiar 
type I form:
\begin{equation}
\label{type1qbc}
\dif W = \left(-\frac{2\pi}{a_o}\right)(8\pi\gamma)(\pb{\gSigma}\cdot r) 
\end{equation}
(see \cite{abk} and \cite{ack}).

\section{Quantization and entropy}

The situation we see now is formally identical to the
situation encountered in the type I calculation: we have a $U(1)$ 
connection $W$ describing the connection degrees of freedom at the 
horizon, and a surface term in the symplectic structure
which is identical to the Chern-Simons surface term 
appearing in the type I case.
Furthermore, the horizon boundary condition in terms of $W$ is the 
same boundary condition that appeared in the type I calculation.

We are therefore
led to the same quantization scheme used in the type I case.
Let us review the scheme.  One first separates the phase space into bulk and 
surface phase spaces, and quantizes each separately -- the bulk using standard 
loop quantum gravity techniques, and quantizing the surface theory as a 
Chern-Simons theory.  We then tensor product the two resulting Hilbert spaces 
together, and impose the quantum version of boundary condition
(\ref{type1qbc}).

Finally we impose the constraints of GR to obtain the final physical Hilbert 
space.

As was done in the type I case, we define the relevant ensemble to
include all horizon area eigenstates with area eigenvalue equal to $a_0$
plus or minus some tolerance $\delta$, all of these eigenstates being
equally weighted. Since the ensemble is the same as in the type I
calculation, the entropy is in fact the same.
Thus, we again reproduce the Bekenstein Hawking entropy
using the same value of the Barbero-Immirzi parameter
used in the type I case.

\section{Type II geometry operators}

But if we wish to be more ambitious, gain a better intuition for the
horizon geometry in terms of its type II character, or at least gain a 
more complete characterization of the ensemble just constructed, it is
important to construct operators corresponding to quantities more
directly related to the type II nature of the horizon.  Let us begin
by building operators at the level of the kinematical Hilbert space.
At that level of the quantization we have the following picture.
\begin{figure}
  \begin{center}
  \includegraphics[height=4cm]{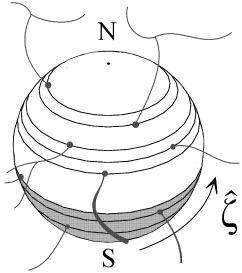}
  \caption{Behavior of $\hat{\zeta}$ eigenvalues}\label{zetahatfig}
  \end{center}
\end{figure}
The kinematical Hilbert space is spanned
by spin-network states.  Each spin-network state may be visualized
as a graph embedded in $3$-space, with verticies allowed to lie
on the inner $2$-sphere boundary $S$.  These vertices are referred 
to as ``punctures."  The
punctures are both the sources for the surface Chern-Simons theory, and the 
locations where horizon surface area is ``concentrated" in discrete amounts.  
This is the physical picture.

In the present, type II case, we have a further basic structure
entering the picture.  At the classical level, given any 
$S^2$ cross-section $S$ of a type II isolated 
horizon with sufficiently generic multipoles, the axial symmetry field 
Lie-dragging the intrinsic geometry of $S$ is unique.  The orbits of this 
symmetry field give us a foliation of $S$ into circular leaves; 
we call this foliation ``axial" and denote it by ``$\xi$".  
$\xi$ is a pure gauge degree of freedom on the inner $2$-sphere boundary.  
That is, the group of diffeomorphisms acts transitively on the space of 
possible $\xi$'s.

To extract the physics of the situation in the quantum theory,
we simply fix $\xi$. This may be thought of as ``gauge-fixing."

Regarding the legitimacy of this:  It would be more 
satisfactory if the gauge-fixing could be done as true gauge fixing.  The 
problem is that if one \textit{actually} gauge fixes $\xi$ to be equal to some 
background $\xi_0$, that reduces the group of diffeomorphisms we divide out by 
at the quantum level.  This reduction of the gauge group causes problems 
when we take into consideration the handling of certain extra structures 
necessary in the quantization of the surface phase space. In fact, the final 
entropy we calculate if we do this is ambiguous.  Our viewpoint is that this 
problem is due to the fact that, by fixing $\xi$, we have broken 
diffeomorphism invariance, and diffeomorphism invariance is ``sacred."  

Nevertheless, we need a fixed $\xi$ to build certain 
physical operators --- so we fix one.  There is no harm in doing
this because in the physical Hilbert space one divides out by 
diffeomorphisms anyway: the choice of a fixed $\xi$ does not
matter at that level.
The most important operators we will build using $\xi$ are the 
multipole operators, operators which will carry over to the 
physical Hilbert space.

For convenience, let us furthermore introduce a coordinate $\zeta_0$
labelling the leaves of $\xi$.  Nothing we do is going to 
depend on the choice of this coordinate; it is introduced
merely for convenience.

Next, we introduce an \textit{operator} corresponding to the preferred 
coordinate $\zeta$ introduced classically in (\ref{stdform}).  
Classically, the $\zeta$ coordinate
has the convenient property that it increases from South to North in 
proportion to area:  
\begin{figure}
  \begin{center}
  \includegraphics[height=4cm]{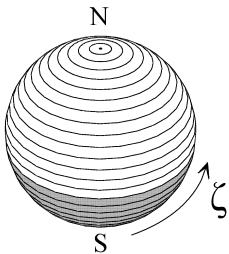}
  \caption{Classical $\zeta$ increases in proportion to area}\label{zetafig}
  \end{center}
\end{figure}
\begin{equation}
\label{zetacl}
\zeta(\zeta_0') = -1 + \frac{2 a_{\zeta_0 < \zeta_0'}}{a_S}
\end{equation} 
So that we define
\begin{equation}
\label{zetaquant}
\hat{\zeta}(\zeta_0') = -1 + \frac{2 \hat{a}_{\zeta_0 < \zeta_0'}}{\hat{a}_S}
\end{equation} 
where $\hat{a}_{\zeta_0 < \zeta_0'}$ is the area operator corresponding to the 
portion of $S$ defined by $\zeta_0 < \zeta_0'$.  

As $\hat{\zeta}$ is an operator-valued
function on the sphere, its eigenvalues may be thought of as
functions on the sphere.

Let us gain a picture of the behavior of these eigenvalues.
Given a spin-network state, recall area is concentrated at the
punctures in discrete amounts.  Consequently, the eigenvalues of $\hat{\zeta}$ 
jump discontinuously at leaves which contain
punctures and everywhere else the eigenvalues are constant.  
This is represented in figure (\ref{zetahatfig}).  

Next, we define an operator for $\Psi_2$. On the original classical
phase space, the multipoles are fixed.  One can show this means that 
$\Psi_2$ \textit{as a function of $\zeta$} is completely fixed.
Specifically, if we set $4\pi R_0^2 := a_o$, the function is given by
\begin{equation}
\mathring{\Psi}_2(x)
= -\frac{1}{R_0^2} \sum_n (\Io_n+i\Lo_n)Y_{n,0}(x),
\quad \left(\mathring{\Psi}_2:[-1,1]\rightarrow \C \right)
\end{equation}
That is, on the classical phase space, $\Psi_2 = \mathring{\Psi}_2(\zeta)$.  An
obvious definition is then
\begin{equation}
\hat{\Psi}_2 := \mathring{\Psi}_2 (\hat{\zeta})
\end{equation}

Finally, we define the multipole operators.  The basic definition
for the multipoles is given by 
\begin{equation}
I_n+iL_n = -\oint_S \Psi_2 Y_{n,0}(\zeta)\areaform
= - \frac{a_S}{2}\int_{-1}^1\Psi_2 Y_{n,0}(\zeta)\dif \zeta
\end{equation}
One would like to simply take this
expression directly over to the quantum theory.  However, as it stands, 
there is a problem with the integrand.
Because the $\hat{\zeta}$ eigenvalues depend on position in a discontinous
manner at the punctures, the eigenvalues of $\dif \hat{\zeta}$ will have 
$\delta$-functions at the punctures.  But the other
elements of the integrand, as they are functions of $\hat{\zeta}$,
will be \textit{discontinuous} at the punctures.  Consequently, the 
meaning of the expression is ambiguous: we have delta functions
multiplied into discontinous functions. 

The way we choose to regularize is simply to 
replace the eigenvalues $\zeta$, which are discontinuous, with
a family of smooth $\zeta_i$'s that converge to the 
physical $\zeta$ in the limit $i \rightarrow \infty$.  We then take the
limit $i \rightarrow \infty$:
\begin{eqnarray}
\nonumber
\hat{I}_n+i\hat{L}_n
&=& -\lim_{i\to\infty}\frac{\hat{a}_S}{2}
\int_{-1}^{1}\mathring{\Psi}_2(\hat{\zeta}_i) 
Y_{n,0}(\hat{\zeta}_i)\dif \hat{\zeta}_i
\\
\label{multops}
&=& \frac{\hat{a}_S}{a_o}\left(\Io_n+i\Lo_n\right)
\end{eqnarray}
giving us the final expression for the multipole operators.

From this expression, it is easy to see, for example, that
for the ensemble defined in the previous section the relative 
fluctuations in the multipoles will be equal to the relative fluctuations 
in the area. Furthermore, just as $\langle \hat{a}_S \rangle$ is 
within a fixed, small bound ($\delta$) of $a_0$, 
$\langle \hat{I}_n+i\hat{L}_n \rangle$ is within a fixed, small bound of 
$\Io_n+i\Lo_n$ for each n.  Thus we obtain a more complete characterization
of the ensemble and the nature of the fluctuations being allowed.

\section{Summary}

We have mapped the entropy calculation problem for the 
type II case to the type I case via an appropriate choice of 
variable $W$.  In terms of this variable, the surface term
in the symplectic structure is just Chern-Simons, 
and the relation between $W$ and the bulk variables
is the same as in the type I case -- that is, 
the appropriate ``quantum boundary condition" we impose at
the quantum level is still the same. 
Therefore we are able to do the quantization in the same manner 
as was done in the type I case, and get the same entropy.

The difference between the type I and type II cases lies
in the physical interpretation of $W$.  In the type I case, 
the concentrations of $\dif W$ at punctures can be interpreted in
terms of deficit angles (see \cite{abk}).  
In the type II case, on the other hand,
in order to obtain a physical interpretation, we introduced the
$\hat{\zeta}$ operator, $\hat{\Psi}_2$ operator, and multipole operators.

It is worthwhile to note how much of an extension the present work represents.
The space of type II isolated horizons is infinite dimensional, encompassing 
the (finite dimensional) family of Kerr type isolated horizons as well as all 
possible distortions of such horizons compatible with axisymmetry.  A vast 
range of astrophysically realistic black holes are therefore covered.

\section*{Acknowledgements}

This talk was based on joint work done with 
Abhay Ashtekar and Chris Van Den Broeck.  The research was 
supported in part by two Frymoyer Fellowships of Penn State,
the National Science Foundation grant 
PHY-0090091, the Eberly research funds of Penn State and 
the Alexander von Humboldt Foundation of Germany.

\end{document}